\documentclass{PoS}

\title{Overview of Nuclear Physics at Jefferson Lab}

\ShortTitle{Jefferson Lab Overview}

\author{\speaker{R. D. McKeown}%\thanks{A footnote may follow.}
\\
        Thomas Jefferson National Accelerator Facility\\
        E-mail: \email{bmck@jlab.org}}

\abstract{ The Continuous
Electron Beam Accelerator Facility (CEBAF) and associated experimental equipment at Jefferson Lab comprise a unique facility for experimental nuclear physics. This facility is presently being upgraded, which will enable a new experimental program with substantial discovery potential to address important topics in nuclear, hadronic, and electroweak physics. Further in the future, it is envisioned that the Laboratory will evolve into an electron-ion colliding beam facility.}

\FullConference{The 7th International Workshop on Chiral Dynamics,\\
		August 6 -10, 2012\\
		Jefferson Lab, Newport News, Virginia, USA}

\begin{document}
\section{Introduction}

Since 1995, the CEBAF facility at Jefferson Laboratory has operated
high-duty factor (continuous) beams of electrons incident on three
experimental halls (denoted A, B, and C), each with a unique set of experimental
equipment. As a result of advances in the performance of
superconducting radiofrequency (SRF) accelerator technology, the
electron beam has exceeded the original 4~GeV energy specification,
and beams with energies up to 6~GeV with currents up to 180~$\mu$A
have been delivered for the experimental program. In addition, the
development of advanced GaAs photoemission sources has enabled high
quality polarized beam with polarizations up to 85\%. The facility
serves an international scientific user community of over 1300
scientists, and to date over 160 experiments have been completed.

The first decade of scientific results from Jefferson Lab has recently been summarized in an extensive review \cite{decade}. Major achievements include the demonstration of isoscalar meson-exchange effects in the tensor analyzing power in electron-deuteron scattering, the discovery of the unexpected behavior of the ratio of proton electromagnetic form factors $G_E / G_M$ at high momentum transfers, the precise measurements of parity-violating asymmetries that constrain the strangeness content of the form factors to be remarkably small, and pioneering measurements of Generalized Parton Distributions (GPD) in Deeply Virtual Compton Scattering (DVCS) and Transverse Momentum Dependent (TMD) distributions (in Semi-inclusive Deep Inelastic Scattering, or SIDIS). This remarkable record of scientific productivity, and the prospects for further scientific advances in this field,  have motivated a major upgrade of CEBAF to 12 GeV electron beam energy along with substantial new experimental equipment.

\section{12 GeV Upgrade}

Early in the history of CEBAF, plans were developed to
upgrade the capability of the accelerator to enable beams up to 12~GeV
in energy. The US nuclear physics community subsequently endorsed this concept, and the 2002 Long Range Plan of the US Nuclear Science
Advisory Committee (NSAC) \cite{lrp02} contained a
recommendation: ``We strongly recommend the construction of CEBAF at
Jefferson Laboratory to 12 GeV as quickly as possible''. In March
2004 the US Department of Energy (DOE) granted CD-0 approval to
develop a conceptual design for such a facility. Following further development of the
design of the upgraded facility, the 2007 NSAC
Long Range Plan \cite{lrp07} reaffirmed the community commitment to
this project: ``We recommend completion of the 12 GeV Upgrade at
Jefferson Lab.  The Upgrade will enable new insights into the
structure of the nucleon, the transition between the hadronic and
quark/gluon descriptions of nuclei, and the nature of confinement.''
The Jefferson Lab user community and the
Laboratory staff completed the design, including detailed cost and schedule, in 2008. DOE then approved the start of
construction in September 2008.

CEBAF is currently a recirculating linac, with 2 linac sections, each
consisting of 20 cryomodules. A cryomodule contains 8
superconducting RF cavities and is capable of
an average 25~MeV of acceleration. Thus each linac section is nominally capable of producing 0.5~GeV
of energy gain. The recirculating arcs contain
quadrupole and dipole magnets in separate beamlines that facilitate
beam acceleration up to 5 times through both linacs, producing
a nominal energy of 5~GeV with actual performance up to 6~GeV. Individual beam pulses can be "kicked" into an
extraction line after the second (South) linac section and delivered into one of the 3 experimental halls. Thus the
beam can be split into three simultaneous 499~MHz beams with
energies in multiples of $\frac{1}{5}$ of the full 5-pass energy. The high duty factor associated with the 2~ns beam structure
enables high luminosity experiments with coincident detection of multiple particles per event.

\begin{figure}
%\begin{center}
\includegraphics[width=7in]{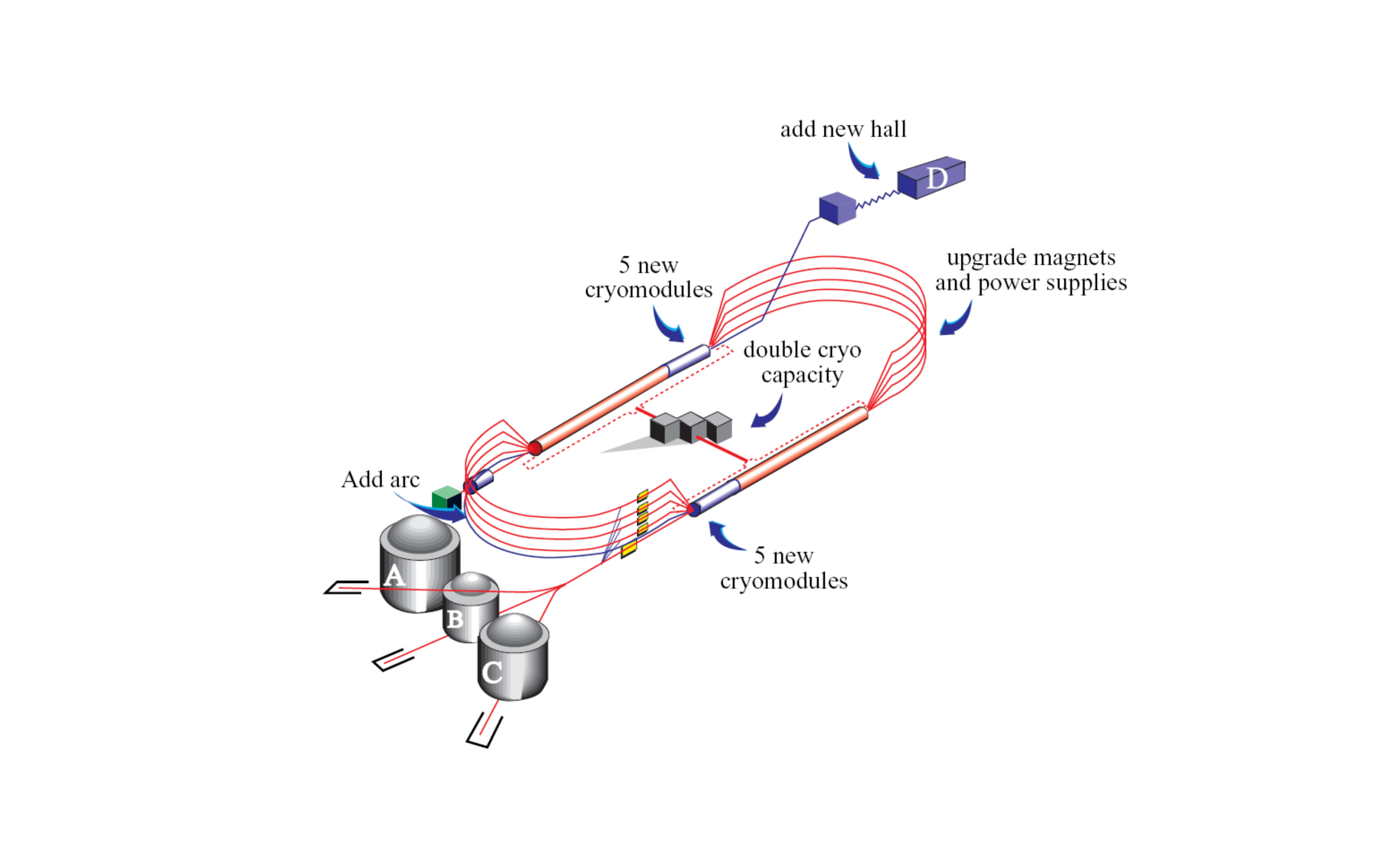}
%\end{center}
\caption{\label{fig:overview}Jefferson Lab 12 GeV upgrade concept.}
\end{figure}

Fig.~\ref{fig:overview} illustrates the basic concept of the 12~GeV upgrade project.
In addition to the upgrade of the
accelerator system to enable delivery of 12~GeV beam, the
experimental equipment will be enhanced to facilitate full
exploitation of the higher energy beam. This includes substantial
new equipment in Hall B and Hall C, and a completely new Hall D with
a new detector and spectrometer system. The plan for Hall A also
includes upgraded and new equipment that is outside the present
construction project.

\section{12 GeV Science Program}

The physics program to be addressed with the Jefferson Lab 12~GeV upgrade has been developed in collaboration with the user
community and with the guidance of the Program Advisory Committee. There are presently 52 approved experiments, and 15
additional proposals have conditional approval.

The major science questions to be addressed with the upgraded facility include:
\begin{itemize}
  \item What is the role of gluonic excitations in the spectroscopy of light mesons? Can these excitations elucidate the origin of quark confinement?
  \item Where is the missing spin in the nucleon? Is there a significant contribution from valence quark orbital angular momentum?
  \item Can we reveal a novel landscape of nucleon substructure through measurements of new multidimensional distribution functions?
  \item What is the relation between short-range N-N correlations, the partonic structure of nuclei, and the origin of the nuclear force?
  \item Can we discover evidence for physics beyond the standard model of particle physics?
\end{itemize}

A detailed exposition of the science opportunities at the upgraded facility has recently been published \cite{12WP}. Here we briefly present some highlights.

\section{Meson Spectroscopy}

 \begin{figure}
%\begin{center}
\includegraphics[width=5in]{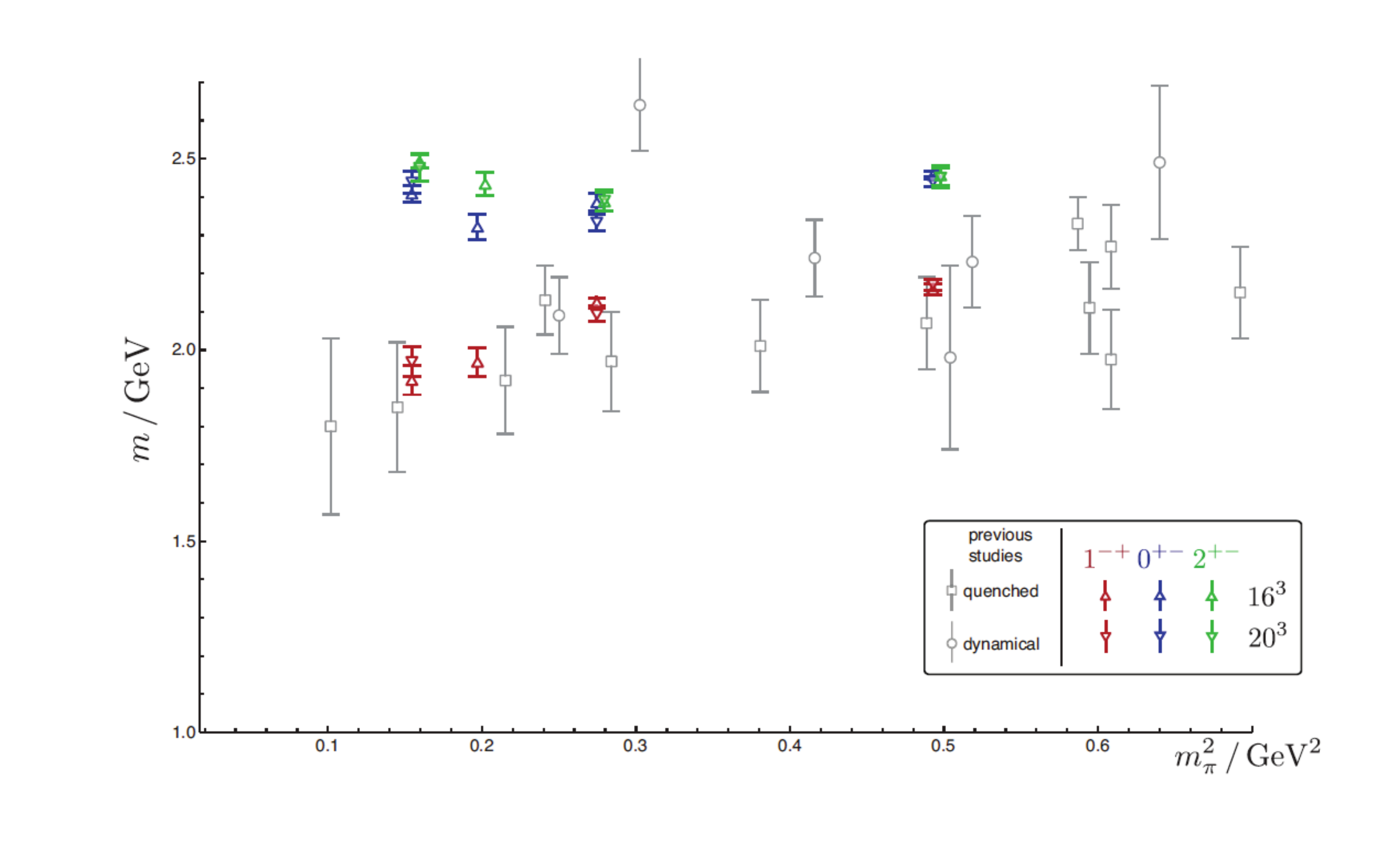}
%\end{center}
%\vspace{-2.5in}
\caption{\label{fig:Mesons} Lattice QCD predictions for exotic isovector mesons \cite{LQCD2}.}
\end{figure}

 Excitation of the gluonic field in a quark-antiquark system can lead to mesonic states with exotic quantum numbers
 ($J^{PC}= 0^{+-}, 1^{-+}, 2^{+-}$) that cannot be
described by states with only quark-antiquark degrees of freedom. These states and their
properties have recently been studied in detail using lattice QCD methods
\cite{LQCD, LQCD2}. As shown in Fig.~\ref{fig:Mesons}, chiral extrapolation of these lattice calculations indicate that the exotic mesons will indeed be present in the mass range 2-2.5~GeV.

\begin{figure}
%\begin{center}
\includegraphics[width=6in]{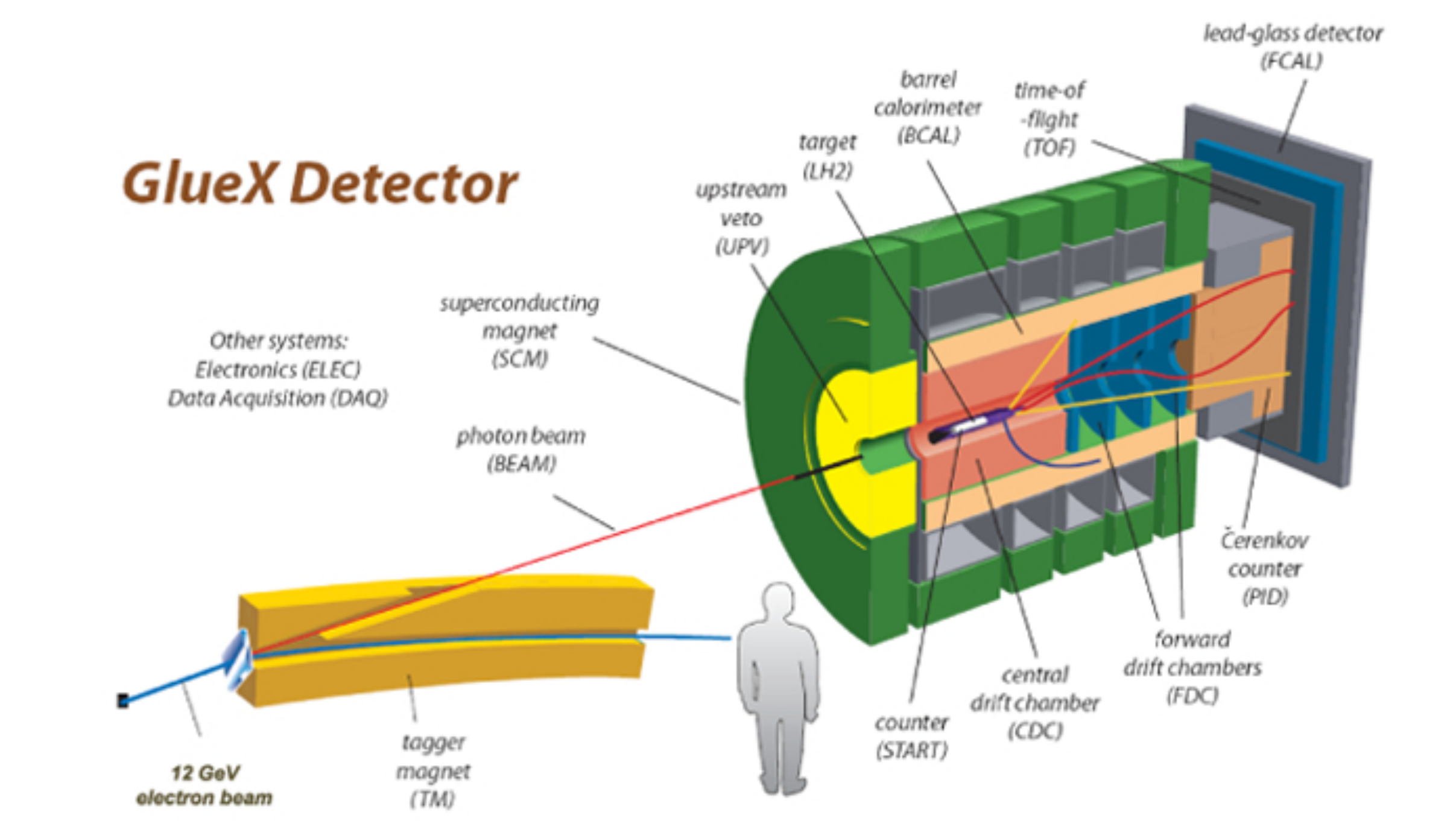}
%\end{center}
\vspace{-.25in}
\caption{\label{fig:gluex}Schematic of GlueX apparatus.}
\end{figure}

A major new experiment, known as GlueX
(see Fig.~\ref{fig:gluex}), will be constructed and sited in Hall D. The main goal of the
GlueX experiment is to search for exotic mesons produced via
photoproduction on the nucleon.  Linearly polarized photons will be produced upstream of Hall D by a
coherent bremmstrahlung process using a thin ($\sim 20$ micron) diamond wafer.
The scattered electrons from 8.5-9~GeV bremmstrahlung photons will be tagged with
scintillator detectors following a bending magnet, yielding a tagged
photon resolution of 0.2\% with fluxes expected to reach $10^8$/s.

\section{Nucleon Structure}

Since the initial discovery at SLAC in the 1960's, the partonic structure of the nucleon has been described by the one-dimensional parton distribution functions (PDF). The four parity conserving structure functions of the variable $x$ ($f_1$, $f_2$, $g_1$, and $g_2$), have been studied in great detail in many experiments over the last four decades. However, over the last 15 years it has been realized that these one dimensional distributions are not adequate to describe the nucleon and do not contain some essential physics needed to provide a complete picture. The orbital angular momentum of partons is a good example of a degree of freedom that is not exhibited in the standard 1-d PDFs. Recent experimental and theoretical studies indicate that a more complete description of the partonic structure of the nucleon is realizable through new multi-dimensional distributions: Generalized Parton Distributions (GPD) and Transverse Momentum Dependent (TMD) distributions.

\begin{figure}
\begin{center}
\includegraphics[width=4in]{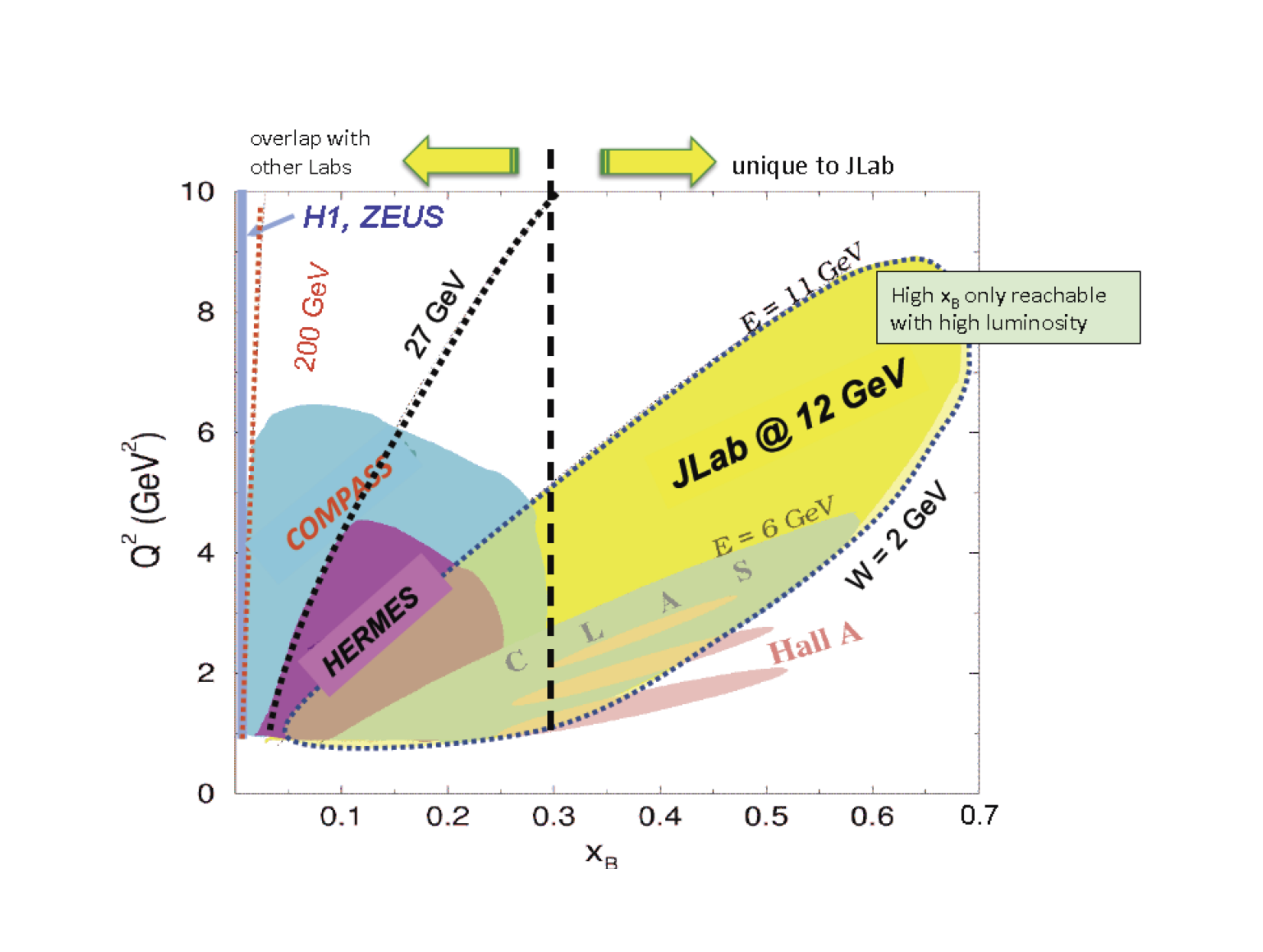}
\end{center}
%\vspace{-2.5in}
\caption{\label{fig:Q2_x}Kinematic reach of the 12 GeV upgrade at Jefferson Lab in $Q^2$ and $x$ space, in relation to other previous facilities and experiments \cite{12WP}. The higher energy, combined with the higher luminosity available at JLab, offers new opportunities to explore nucleon structure in the valence quark regime. }
\end{figure}

GPDs contain information on the correlation between the quark/gluon transverse position in the nucleon and its longitudinal momentum They can be accessed in exclusive scattering processes at large $Q^2$: deeply virtual Compton scattering (DVCS) and deep virtual meson production (DVMP).  GPDs offer a path to a full 3-dimensional exploration of nucleon structure, in transverse position and longitudinal momentum space, enabling spatial tomography of the nucleon. The new CLAS12 apparatus being constructed for Hall B as part of the 12 GeV upgrade is specifically designed to study these processes.

Transverse momentum dependent distributions (TMDs) contain information on the quark/gluon intrinsic motion in a nucleon, and on the correlations between the transverse momentum of the quark and the quark/nucleon spins.  TMDs offer a unique opportunity for a momentum tomography of the nucleon, and can be measured in Semi-Inclusive Deep Inelastic Scattering (SIDIS), in which the nucleon is no longer intact and one of the outgoing hadrons is detected. SIDIS will be studied in Hall C with high resolution spectrometers, in Hall B with CLAS12, and in Hall A with the new Super Bigbite Spectrometer (SBS). A major new capability for SIDIS is proposed in the Solenoidal Large Intensity Device (SoLID) to be sited in Hall A. This project is still under development, but would offer exceptional capability for mapping out the TMDs with the 12 GeV beam at JLab.

As shown in Fig.~\ref{fig:Q2_x}, the upgraded facility at Jefferson Lab will significantly extend the kinematic reach for deep-inelastic studies of the nucleon in the valence region. This new coverage will enable a complete exploration of SIDIS and also inclusive scattering to provide a more complete picture of the nucleon in the valence region.

\section{Quarks in Nuclei}

The Jefferson Lab 12 GeV Upgrade will both study the QCD structure of nuclei and use the nucleus as a laboratory to study QCD.  The new facility will enable investigations of a number of the most fundamental questions in modern nuclear physics.

The nature of the nucleon-nucleon (NN) relative wave function at short distances is fundamental to the origin of the nuclear force and to the properties of nuclei. It is not known if this system can be described only in terms of nucleons and mesons, or whether quarks and gluons are necessary for its description. Recent studies indicate that modification of the nuclear parton distributions, or the "EMC effect", is related to short-range NN correlations in nuclei \cite{EMC_SRC}. Further experimental studies to explore these important issues will be possible with the 12 GeV CEBAF and new experimental equipment.

QCD also suggests the existence of novel phenomena in nuclear physics.  The nuclear medium provides mechanisms for filtering quantum states and studying their spacetime evolution. Studying the hadronization of a struck quark in different nuclei affords a unique method for elucidating this process. The formation of small color singlet configurations leads to the novel process known as color transparency. The increased kinematic range of the 12 GeV CEBAF will offer new opportunities to study these and other related topics.

A recent Jefferson Lab experiment, PREX, has demonstrated a new method to determine the neutron radius of a heavy nucleus like $^{208}$Pb \cite{PREX}. The precise measurements of the charge distribution of nuclei in elastic electron scattering provide stringent constraints on the distribution of protons in nuclei. However, the distribution of neutrons is more difficult to study and is also quite important for predicting the properties of neutron stars. The weak charge of the neutron is -1, whereas the weak charge of the proton is $1-4\sin^2 \theta_W \ll 1$. Thus, the measurement of parity violating asymmetries in elastic electron scattering from nuclei is sensitive to the neutron distribution and can be used to constrain the neutron radius. Future studies with higher precision are planned for the Jefferson Lab 12 GeV program.

\section{Beyond the Standard Model}

Exciting new opportunities to search for new physics beyond the Standard Model will become possible at Jefferson Lab in the 12 GeV era. The very precise measurements of parity violating asymmetries to study the strange form factors in elastic electron-proton scattering have demonstrated that this technique has substantial potential for precision tests of the Standard Model. The strength of the neutral weak interaction is parameterized in the standard model by the weak mixing angle $\theta_W$. This parameter is very precisely determined at the $Z$ boson mass by $e^+$-$e^-$ collider experiments. The two best measurements (which differ by more than $2\sigma$) have uncertainties of 0.00029 and 0.00026, and can be combined to yield the average value $\sin^2 \theta_W =  0.23116  \pm 0.00013 $ \cite{pdg}. Radiative corrections associated with standard model physics predicts a ``running'' of this coupling to $\sin^2 \theta_W =  0.2388$ at $Q^2 =0$. Additional particles at high mass (larger than $M_Z$) would generally modify these radiative corrections, leading to a different value of $\sin^2 \theta_W $ at $Q^2=0$. Thus precise measurements of the neutral weak interaction at low $Q^2 \ll M_Z^2$ can reveal the presence of particles and forces not present in the standard model.

A major new experiment to study parity-violation in elastic electron-proton scattering at low $Q^2$ was recently completed in Hall C at Jefferson Lab\cite{Qweak}.
The data from this experiment, known as $Q_{weak}$, are under analysis. In addition,
there are presently 2 new proposals to perform parity violation measurements at the
upgraded CEBAF. One would use the proposed solenoidal magnetic
spectrometer system (SOLID) to study parity-violating deep inelastic
scattering \cite{SOLID}. The other proposal involves the
construction of a novel dedicated toroidal spectrometer to study
parity-violating M{\o}ller scattering \cite{Moller}. Both
experiments will require construction of substantial new
experimental equipment (beyond the scope of the present upgrade
project) and are proposed to be sited in experimental Hall A.

\begin{figure}
\begin{center}
\includegraphics[width=4.5 in]{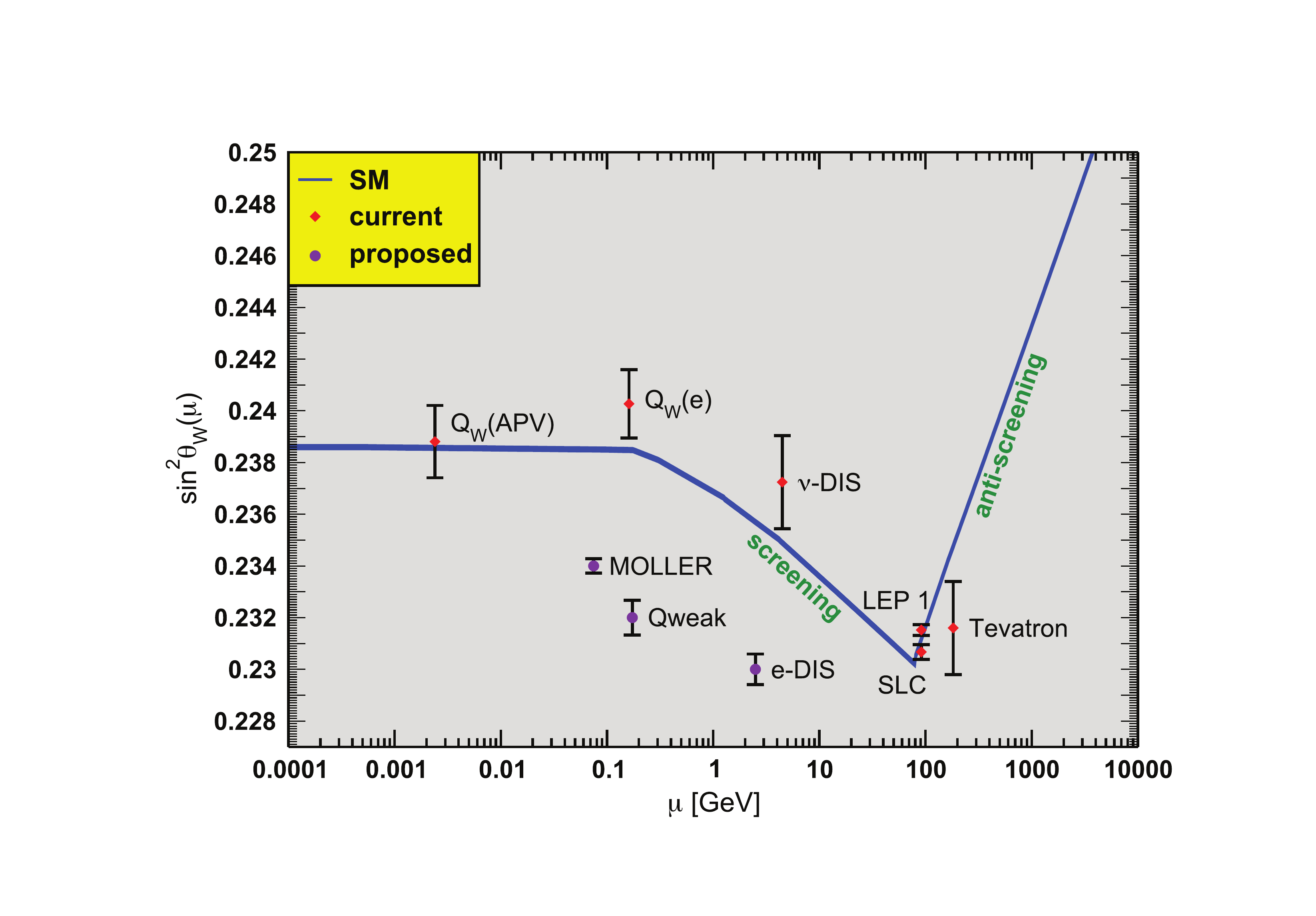}
\end{center}
\caption{\label{fig:PV} The standard model prediction for $\sin^2 \theta_W$, along with previous measurements and expected results for future experiments at Jefferson Lab \cite{pdg}.}
\end{figure}

Fig.~\ref{fig:PV} shows the standard model prediction (in the "minimal subtraction", or $\overline{\rm MS}$, renormalization scheme) for $\sin^2 \theta_W$ as a function of energy scale $\mu$. The value at the mass of the $Z$ boson is fit to the $e^+-e^-$ data. Also shown are results from atomic parity violation, parity violating M{\o}ller scattering at SLAC (E158), and results from deep inelastic neutrino scattering.  (It should be noted that the nuclear corrections for the deep inelastic neutrino scattering results are still a subject of substantial discussion.) The projected results
for $\sin^2 \theta_W$ for the future Jefferson Lab experiments are shown at the correct energy scale, but arbitrary values of $\sin^2 \theta_W$, to illustrate the expected experimental uncertainty.

Heavy photons, called A's, are new hypothesized massive vector bosons that have a small coupling to electrically charged matter, including electrons.  The existence of an A' is theoretically natural and could explain the discrepancy between the measured and observed anomalous magnetic moment of the muon \cite{g-2} and several intriguing dark matter-related anomalies.  New electron fixed-target experiments proposed at Jefferson Lab, with its high-quality and high-luminosity electron beams, present a unique and powerful probe for A's.  These experiments include the A' Experiment (APEX)\cite{APEX}, the Heavy Photon Search (HPS)\cite{HPS}, and Detecting A Resonance Kinematically with Electrons Incident on a Gaseous Hydrogen Target (Dark Light)\cite{DARKLIGHT}.

\section{Electron Ion Collider}

The 2007 NSAC Long Range Plan \cite{lrp07} identified an Electron Ion Collider as a new opportunity for the field of nuclear physics: "An Electron-Ion Collider (EIC) with polarized beams has been embraced by the U.S. nuclear science community as embodying the vision for reaching the next QCD frontier.  EIC would provide unique capabilities for the study of QCD well beyond those available at existing facilities worldwide and complementary to those planned for the next generation of accelerators in Europe and Asia." In 2010, a 10 week program at the Institute for Nuclear Theory in Seattle explored the scientific case for such a collider \cite{INT11}. The community reached a consensus on the basic scientific requirements for such a facility:
\begin{itemize}
  \item Highly polarized ($\sim 70$\%) electron and nucleon beams
  \item Ion beams from deuteron to the heaviest nuclei (Uranium or Lead)
  \item Variable center of mass energies from $\sim 20 -  \sim 100$~ GeV, upgradable to $\sim 150$~ GeV
  \item High collision luminosity $\sim 10^{33-34} {\rm cm}^{-2} {\rm s}^{-1}$
\end{itemize}
Subsequently, a white paper has been generated to coherently elucidate the physics that could be addressed with an EIC \cite{EICWP12}. Such a facility would capitalize on the powerful new experimental techniques for exploring nucleon structure that are being developed for the 12 GeV JLab program, and apply them to the low $x$ region where the dynamics is dominated by the gluons. It is widely perceived that addressing this kinematic regime with high luminosity and fully polarized beams is necessary to complete our understanding of the basic partonic structure of the nucleon. An EIC is viewed as a natural extension of the capabilities of the Jefferson Lab 12 GeV upgrade, the RHIC spin program, and the COMPASS experiment at CERN.

\begin{figure}
\begin{center}
\includegraphics[width=4.5 in]{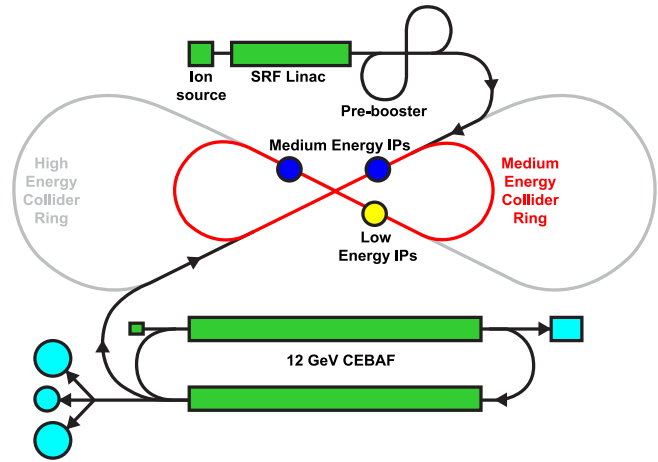}
\end{center}
\caption{\label{fig:MEIC} A schematic drawing of MEIC. The high energy ring is a future possibility.}
\end{figure}

The Accelerator and Physics Divisions at Jefferson Lab have been working for several years on a novel design for an EIC that would utilize the 12 GeV CEBAF as an injector to a collider facility \cite{MEIC12}. As shown in Figure~\ref{fig:MEIC}, the storage rings would be in a "figure 8" layout to mitigate the effects of depolarizing resonances and facilitate high beam polarization. A medium energy version, MEIC, is envisioned that would collide 12 GeV electrons with 100 GeV protons. This could be upgraded to the full EIC facility with 12 GeV electrons colliding with 250 GeV protons.

\section{Conclusion}

Jefferson Lab will continue its tradition of providing forefront nuclear physics capabilities with intense polarized electron beams. The 12 GeV upgrade which is presently in progress will enable dramatic advances in the study of meson spectroscopy, nucleon structure, the short-distance structure of nuclei, and precision tests of the Standard Model. The program that is presently planned will require at least 7-10 years to execute after the facility becomes operational in 2015. Further in the future, it is envisioned that the Lab will evolve to an electron ion collider facility. Such a facility would address a new QCD frontier that would enable new scientific opportunities well into the future beyond the 12 GeV upgrade.

\section{Acknowledgement}

This work was
supported by DOE contract DE-AC05-06OR23177, under which Jefferson
Science Associates, LLC, operates the Thomas Jefferson National
Accelerator Facility.

\end{document}